\documentclass[aps,prl,twocolumn,preprintnumbers,superscriptaddress,amsmath,amssymb]{revtex4-1}
\usepackage[normalem]{ulem}
\usepackage{graphicx}
\usepackage{subfigure}
\usepackage{epsfig}
\usepackage{xcolor}
\usepackage{dcolumn}
\usepackage{bm}
\usepackage{ulem}
\usepackage{color}
\usepackage{amsmath}
\usepackage{amsfonts}
\usepackage{amssymb}

\usepackage{epstopdf}

\begin{document}

\title{Widely tunable band gap in a multivalley semiconductor SnSe by potassium doping}

\author{Kenan Zhang}
\affiliation{State Key Laboratory of Low Dimensional Quantum Physics and Department of Physics, Tsinghua University, Beijing 100084, China}

\author{Ke Deng}
\affiliation{State Key Laboratory of Low Dimensional Quantum Physics and Department of Physics, Tsinghua University, Beijing 100084, China}

\author{Jiaheng Li}
\affiliation{State Key Laboratory of Low Dimensional Quantum Physics and Department of Physics, Tsinghua University, Beijing 100084, China}

\author{Haoxiong Zhang}
\affiliation{State Key Laboratory of Low Dimensional Quantum Physics and Department of Physics, Tsinghua University, Beijing 100084, China}

\author{Wei Yao}
\affiliation{State Key Laboratory of Low Dimensional Quantum Physics and Department of Physics, Tsinghua University, Beijing 100084, China}

\author{Jonathan Denlinger}
\affiliation{Advanced Light Source, Lawrence Berkeley National Laboratory, Berkeley, California 94720, USA}

\author{Yang Wu}
\affiliation{Department of Physics and Tsinghua-Foxconn Nanotechnology Research Center, Tsinghua University, Beijing, 100084, China}

\author{Wenhui Duan}
\affiliation{State Key Laboratory of Low Dimensional Quantum Physics and Department of Physics, Tsinghua University, Beijing 100084, China}
\affiliation{Collaborative Innovation Center of Quantum Matter, Beijing, China}

\author{Shuyun Zhou}
\altaffiliation{Correspondence should be sent to syzhou@mail.tsinghua.edu.cn}
\affiliation{State Key Laboratory of Low Dimensional Quantum Physics and Department of Physics, Tsinghua University, Beijing 100084, China}
\affiliation{Collaborative Innovation Center of Quantum Matter, Beijing, China}

\begin{abstract}

{\bf SnSe, a group IV\text{-}VI monochalcogenide with layered crystal structure similar to black phosphorus, has recently attracted extensive interests due to its excellent thermoelectric properties and potential device applications. Experimental electronic structure of both the valence and conduction bands is critical for understanding the effects of hole versus electron doping on the thermoelectric properties, and to further reveal possible change of the band gap upon doping.  Here, we report the multivalley valence bands with a large effective mass on semiconducting SnSe crystals and reveal single-valley conduction bands through electron doping to provide a complete picture of the thermoelectric physics.  Moreover, by electron doping through potassium deposition, the band gap of SnSe can be widely tuned from 1.2 eV to 0.4 eV, providing new opportunities for tunable electronic and optoelectronic devices.}

\end{abstract}

\maketitle

\section{INTRODUCTION}

Narrow band gap semiconductors of group IV monochalcogenides, such as PbTe and SnTe, are important thermoelectric materials \cite{Heremans2008Enhancement}. In contrast to monochalcogenides with three-dimensional rock salt structure, layered SnSe with highly anisotropic structure has been recently shown to have exceptional thermoelectric properties. In particular, the figure of merit reaches ZT = 2.6 at 923 K for SnSe \cite{Zhao2014Ultralow}, and significant enhancement of ZT was reported for hole doped SnSe in the low-temperature range from 300 K to 773 K \cite{Zhao2016Ultrahigh}.  Such high thermoelectric performance is contributed by multiple factors with both phononic and electronic origins. The low thermal conductivity has been attributed to anharmonic phonon scattering \cite{Li2015Orbitally}, and the key electronic origin is the multivalley of the band structure. In addition to improved thermoelectric properties, such valley properties can also be useful for room temperature valleytronics, which has been recently reported in a similar compound SnS$_2$ \cite{yjss}. In the two dimensional limit, monolayer SnSe is also very attractive for ultrathin device applications. Different from the cubic structure SnSe \cite{Littlewood2010Band, PhysRevX.7.041020}, low-temperature phase SnSe shares a similar crystal structure to phosphorene \cite{Li2014Black, YeBP} - another intriguing 2D material after graphene \cite{Novoselov2004Electric} and transition metal dichalcogenides \cite{Wang2012Electronics, ZhangH2013}, yet with a lower symmetry due to the presence of two types of atoms. Thus monolayer SnSe is expected to exhibit large spin-orbit splitting piezoelectricity, high ionic dielectric screening \cite{Gomes2015Enhanced}, robust ferroelectricity \cite{Fei2016Ferroelectricity} etc.

So far, experimental studies on the electronic structure of SnSe have been focused on the valence bands of metallic or heavily doped samples  \cite{Wang2017Defects,Lu2017Unexpected,I2017Band,yulinchenprb}, while study on the electronic structure of semiconducting SnSe with the advantage to probe the conduction bands has remained missing. Since the thermoelectric performance can be enhanced significantly either upon hole doping \cite{Zhao2016Ultrahigh} or electron doping \cite{Duong2016Achieving} in the low-temperature range, it is important to measure not only the valence band but also the conduction band in order to obtain a complete picture of the thermoelectric properties, in particular how electron or hole doping affects its thermoelectric performance.  Semiconducting SnSe has an advantage for easy access to the conduction bands through electron doping, and to further reveal the possible change of band gap upon doping. Here we report the electronic structure of semiconducting SnSe single crystals grown using Sn-rich flux synthesis.  We present a systematic analysis of the electronic structure to reveal the multivalley physics and effective mass which is key to its thermoelectric properties. Moreover, by surface doping through potassium (K) deposition, we report a widely tunable band gap from 1.2 eV to 0.4 eV. Our results provide not only direct explanations for the electronic origin of its excellent thermoelectric properties but also offer new possibilities for tunable electronic and optoelectronic devices.

\begin{figure*}
\centering
\includegraphics[width=16.8 cm] {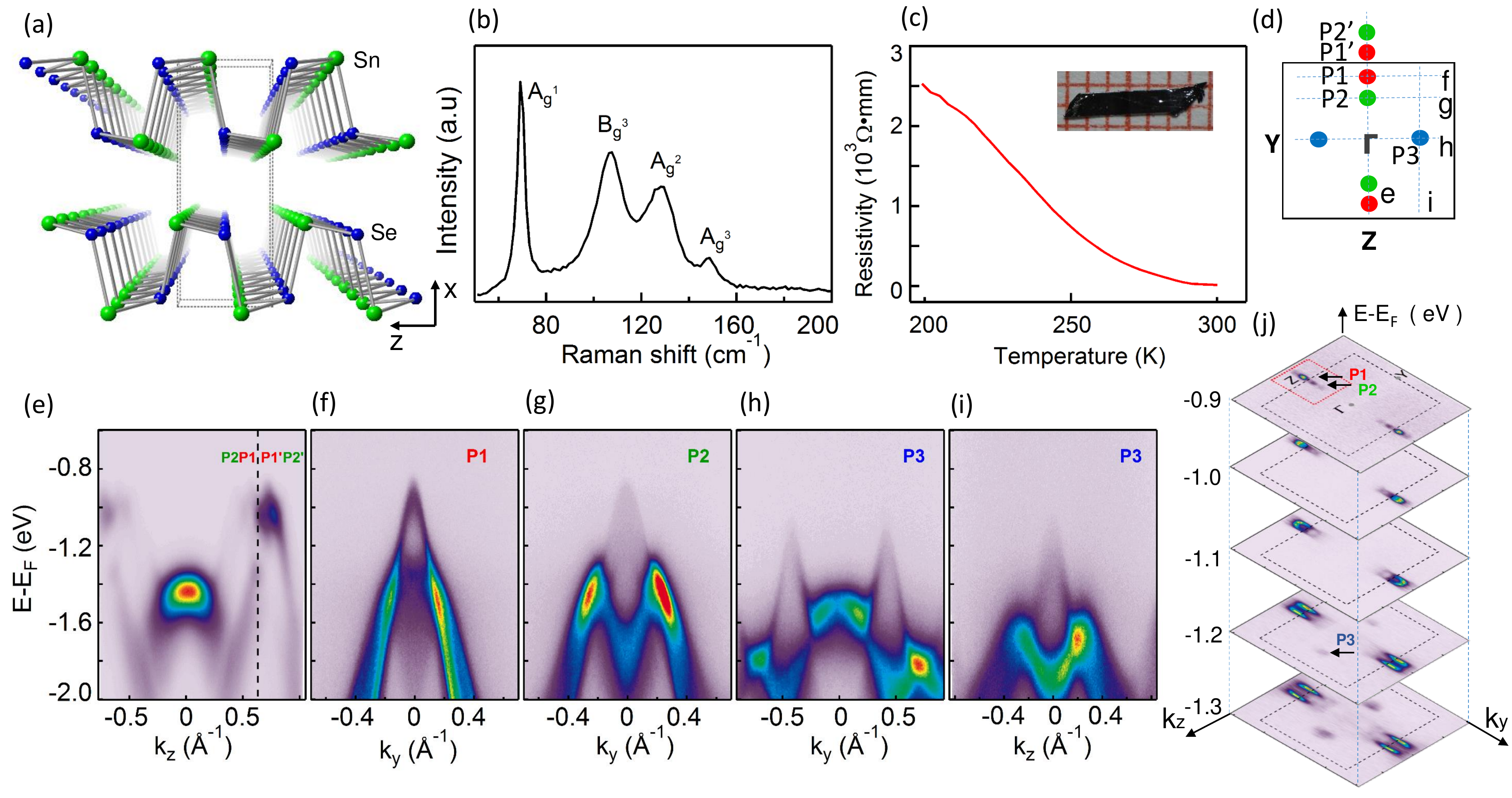}
\label{Figure 1}
\caption{{\bf Crystal and electronic structure of SnSe.} \textbf{(a)} Side view of the crystal structure of SnSe. \textbf{(b)} Raman spectrum measured at room temperature.  \textbf{(c)} Temperature dependent resistivity of SnSe. The inset shows the picture of one single crystal with a few mm size. \textbf{(d)} A schematic drawing of the pockets in the Brillouin zone. \textbf{(e-i)} Different momentum cuts labeled in (d). \textbf{(j)} Intensity maps at constant energies from -0.9 eV to -1.3 eV of SnSe at a photon energy of 35 eV. Black dashed line indicates surface BZ. The gray dots indicate the high-symmetry points of the BZ, and the black arrows indicate the different pockets.}\label{Fig1}
\end{figure*}

\begin{figure*}
\centering
\includegraphics[width=16.8cm] {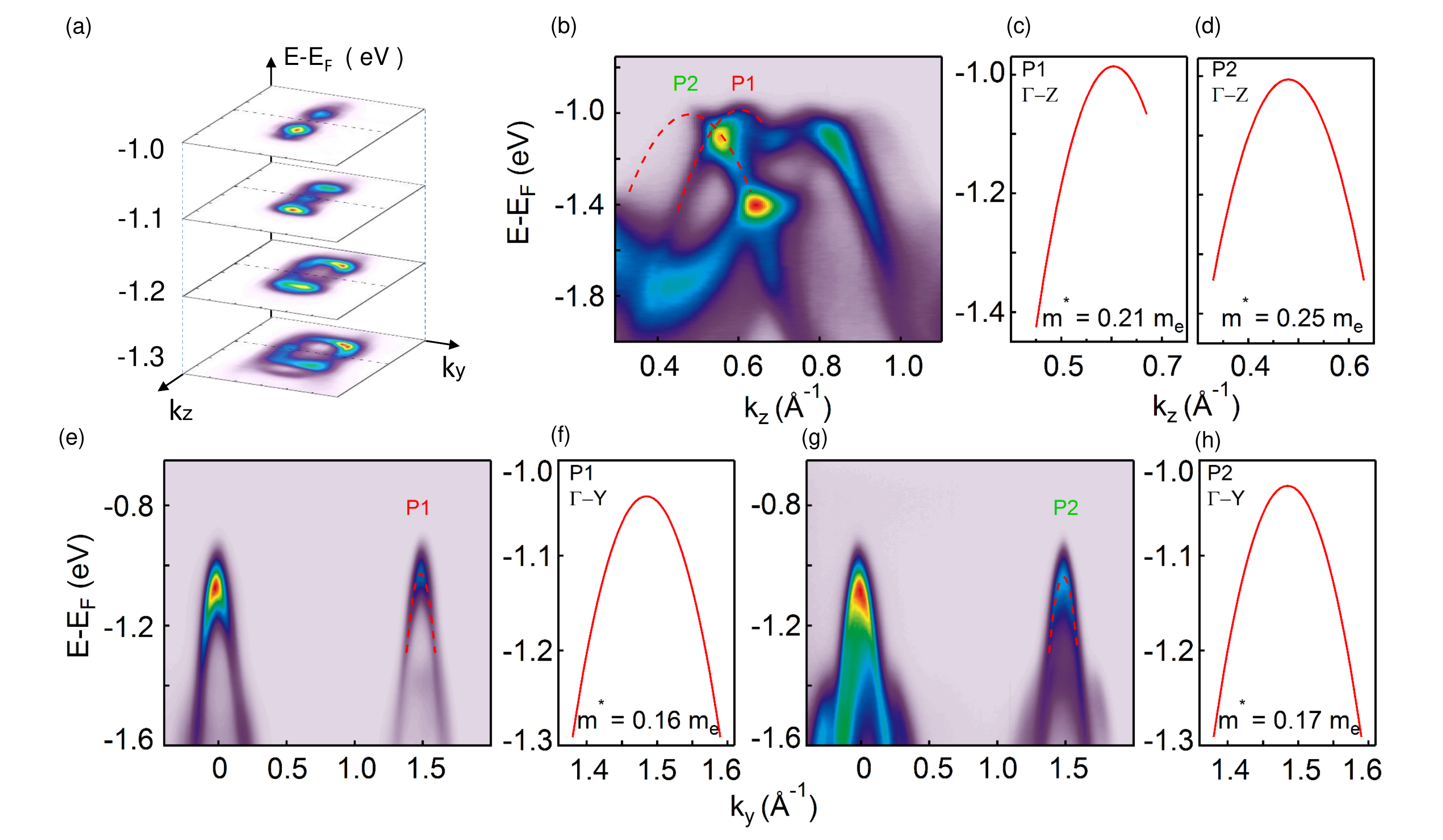}
\label{Fig2}
\caption{{\bf Dispersions and effective mass analysis of the P1 and P2 pockets of SnSe along high symmetry directions.}  \textbf{(a)} Zoom-in of the red dashed rectangular region in Figure \ref{Fig1} (j). \textbf{(b)} Dispersions of pockets P1 and P2 along the $\Gamma$-Z direction. \textbf{(c-d)} Parabolic fitting to extract the effective mass (m$^{*}$) of pockets 1 and 2 along the $\Gamma$-Z direction.  \textbf{(e,g)} Dispersions of pockets P1 and P2 along the $\Gamma$-Y direction, respectively. \textbf{(f,h)} Parabolic fitting line for extracting the effective mass (m$^{*}$) of pockets P1 and P2 along the $\Gamma$-Y direction, respectively. ARPES data in Figure \ref{Fig2} are measured at a photon energy of 80 eV.}\label{Fig2}
\end{figure*}

\begin{figure*}
\centering
\includegraphics[width=16.8 cm] {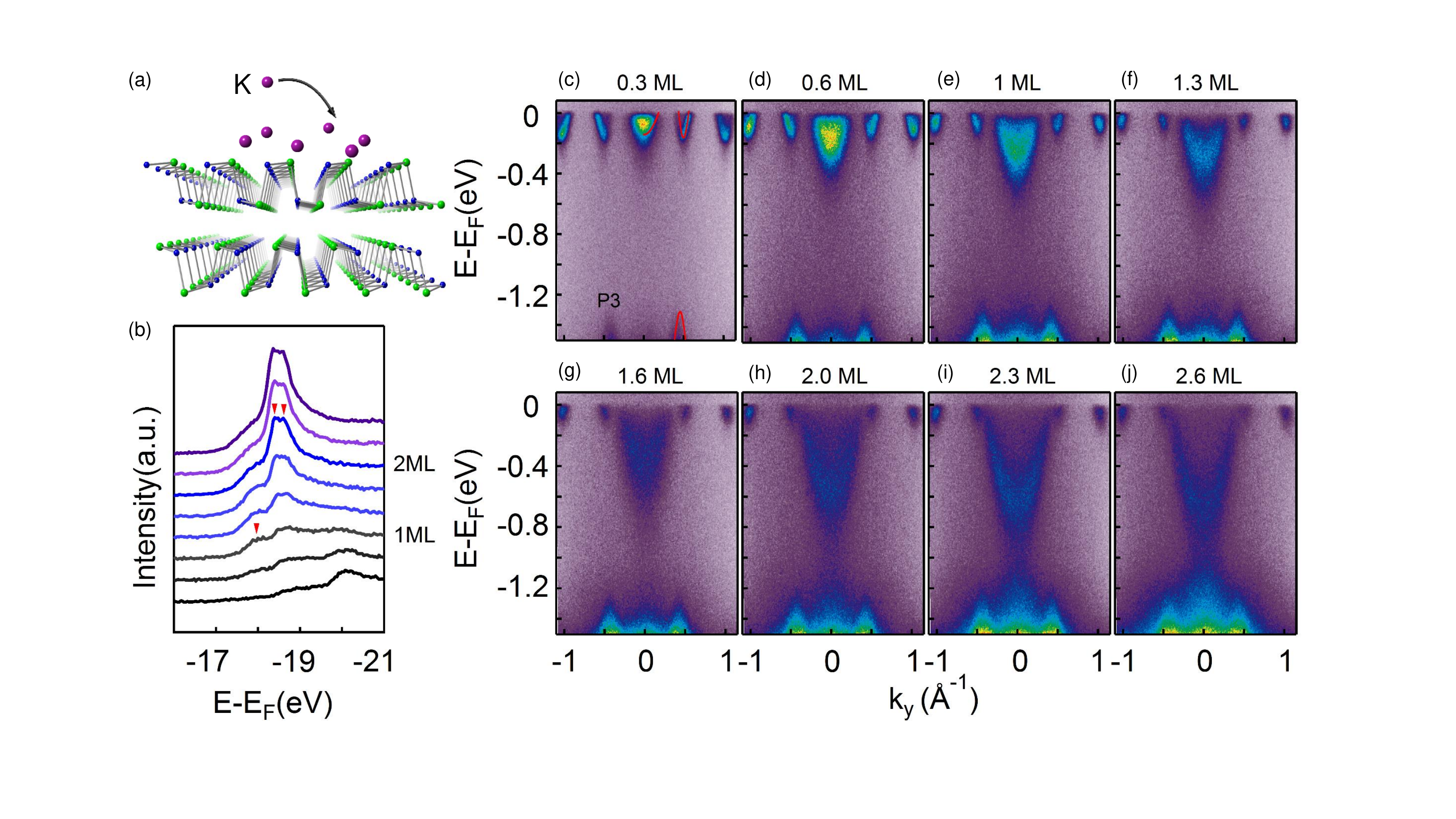}
\label{Fig3}
\caption{{\bf Doping dependence dispersions of SnSe along the $\Gamma$-Y direction.} \textbf{(a)} Schematic illustration of the {\it in situ} surface K deposition. \textbf{(b)} Evolution of K 3p core-level spectra for Figure \ref{Fig3}(c-j), characteristic peaks of K are labbled as red arrows. \textbf{(c-j)} Band structure evolution of SnSe along the $\Gamma$-Y with increasing K doping.  Red solid lines in (c) are calculated dispersions along the $\Gamma$-Y direction  of bulk SnSe for comparison.}\label{Fig3}
\end{figure*}

\begin{figure*}
\centering
\includegraphics[width=16.8 cm] {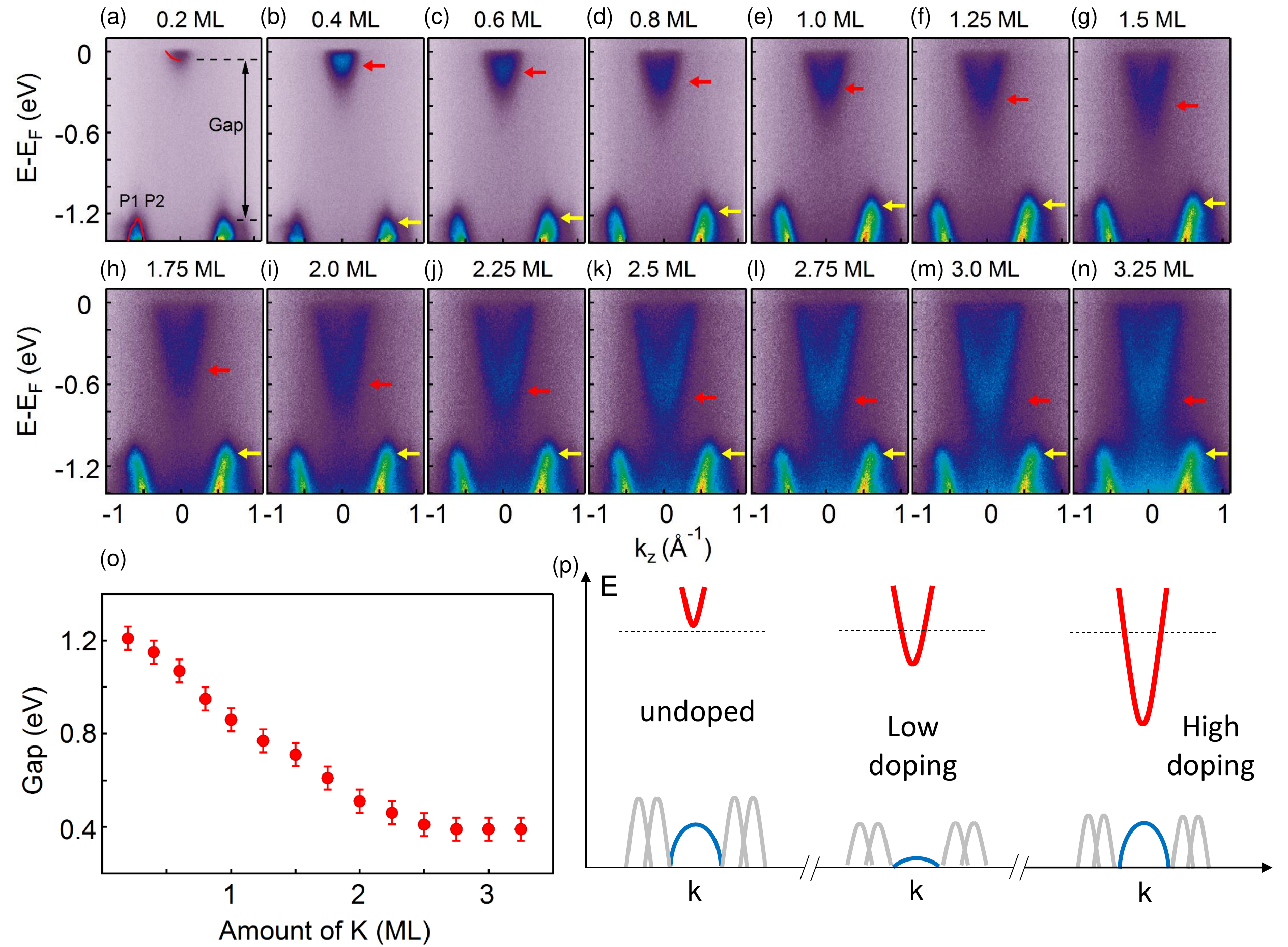}
\label{Fig4}
\caption{{\bf Doping dependence band gap evolution of SnSe.} \textbf{(a-n)} Band structure evolution of SnSe along $\Gamma$-Z with potassium doping at different time. The red and black arrows indicate the CBM and VBM, respectively. Red solid lines are $\Gamma$-Z direction calculated dispersions of bulk SnSe for comparison.  \textbf{(o)} Band gap evolution of SnSe as a function of K dosing. \textbf{(p)} Schematic drawing of the  band evolution with K doping along the $\Gamma$-Y direction.}\label{Fig4}
\end{figure*}

\section{METHODS}

High quality SnSe single crystal was grown by self-flux method. High purity Sn granules (99.99\%, Alfa Aesar) and Se ingot (99.99\%, Alfa Aesar) at a molar ratio of 95:5, were loaded in a silica tube, which is flame-sealed in a vacuum of $\sim$ 1 Pa. The tube was heated at 800$^{\circ}$C for 48 hours to homogenize the starting materials. To minimize the formation of SnSe$_2$ micro-domains, we have used a much slower cooling rate. The reaction was then slowly cooled to 400$^{\circ}$C at 10$^{\circ}$C/h to crystallize SnSe in Sn flux, and the excess Sn was centrifuged isothermally.  The doping concentration of the different batches of samples used in the angle-resolved photoemission spectroscopy (ARPES) measurements are slightly different, resulting in a small variation of Fermi energy by $\pm$ 0.2 eV on different samples. Surface-sensitive ARPES measurements have been performed at BL.4.0.1 and BL.10.0.1 of the Advanced Light Source using photon energies from 30 eV to 80 eV. The crystals were cleaved \textit{in-situ} and measured at a temperature of T$\approx$20 K in vacuum with a base pressure better than 1$\times$10$^{-10}$ Torr. All the first-principle calculations are preformed within the framework of density-functional theory (DFT) by using the plane-wave basis Vienna ab initio simulation package (VASP) code, which implements the Perdew-Burke-Ernzerhof-type generalized gradient approximation. The experimental crystal structure of SnSe bulk is carried out in the whole calculation. A (11$\times$11$\times$5) Monkhorst-Pack grid and a plane-wave energy cutoff of 400 eV is used for self-consistent field calculation. Spin-orbit coupling and Van-der-Waals interaction are also taken into account in our calculations. Since standard DFT calculation usually underestimates the bandgap of materials, the hybrid exchange and correlation functional of Heyd-Scuseria-Ernzerhof (HSE) has been shown to be more successful in predicting energy gaps and other physical properties. A 7$\times$7$\times$3 k-mesh grid is adopted in the HSE calculation.

 \section{RESULTS}

Low-temperature phase SnSe crystallizes in the layered orthorhombic structure with space group Pnma \cite{Chat1986Neutrone}. Figure \ref{Fig1}(a) shows the crystal structure. Each unit cell consists of four formula units and the structure contains two puckered atomically-thick slabs with zigzag and armchair patterns along the b(Y) and c(Z) axes directions, respectively.  Sn and Se atoms within such slabs are covalently bonded, while they are weakly coupled through van der Waals interactions along the a(X) axis. The SnSe structure can also be viewed as a distorted rocksalt structure with Sn-Se-Sn angles deviated from $90^{\circ}$. X-ray and Laue diffractions (see Supplemental Material SI I \cite{Supplemental} for X-ray and Laue diffractions) and Raman measurements reveal the high quality of the single crystals. The Raman spectrum in Fig.~\ref{Fig1}(b) shows four characteristic modes at $\sim$ 69, 108, 128  and 149 cm$^{-1}$ that are assigned to A$_g^1$, B$_g^3$, A$_g^2$ and A$_g^3$ symmetry respectively for the Pnma phase of SnSe \cite{Chandrasekhar1977Infrared,Xu2017In}. The absence of peak at 185 cm$^{-1}$ \cite{Taube2015Temperature} suggests the success of our Sn-rich growth synthetic strategy in avoiding the formation of SnSe$_2$ impurities \cite{Wang2017Defects}. The resistivity in Fig.~\ref{Fig1}(c) shows an increase of resistivity at low temperature, which further verifies the semiconducting property of our SnSe crystals.

The multivalley pockets in the valence band of SnSe are revealed by ARPES measurements. The valence electronic structure of SnSe consists of three hole pockets labeled as P1, P2 and P3 and their locations are schematically shown in Fig.~\ref{Fig1}(d). As these pockets are weakly dispersing along the k$_x$ direction (see Supplemental Material SI II \cite{Supplemental} for photon energy dependence band dispersions), we focus on their dispersions only along the in-plane directions. P1 and P2 are closely spaced, located at (0.00, 0.62$\AA^{-1}$) and  (0.00, $0.47\AA^{-1}$) of the Brillouin zone respectively. Their dispersions along the high symmetric directions are shown in Fig.~\ref{Fig1}(e-g).  The valence band maximum (VBM) of P1 and P2 pockets are nearly equal in energy at $ \approx $ -0.9 eV. This is at much lower energy compared to previous work where the VBM is  located near the Fermi level (E$_F$ ) \cite{Wang2017Defects,Lu2017Unexpected,I2017Band,yulinchenprb}, again supporting the semiconducting property of our samples.  Another hole pocket P3 emerges at slightly lower energy, with the VBM at $ \approx $ -1.2 eV at ($\pm$ 0.42, $0.00\AA^{-1}$). Compared to P1 and P2, P3 has more linear dispersions along both high symmetric directions (see Fig.~\ref{Fig1}(h, i)). Figure \ref{Fig1}(j) shows the ARPES intensity maps measured at energies from -0.9 to -1.3 eV, where the evolution of these hole pockets is clearly revealed.

Since P1 and P2 are directly related to the multivalley physics which is essential for the high thermoelectric coefficient, we show in Fig.~\ref{Fig2} a detailed analysis of these pockets and the effective mass to reveal the electronic origin of the thermoelectric properties.  The evolution of these two pockets with energy is revealed in the zoom-in plot  shown in Fig.~\ref{Fig2}(a).  These pockets, separated at the VBM, gradually grow in size and finally merge together at lower energy.  The effective mass m$^*$ of the band structure at the VBM is a critical factor for the high Seebeck coefficient S and the power factor $\sigma$ for thermoelectric materials. By fitting the dispersion of these two bands near the VBM with a parabolic function (Fig.~2(b-h)), the extracted effective mass along the $\Gamma$-Z and $\Gamma$-Y directions is m$^{*}_{P1_Z}$  = 0.21 m$_e$, m$^{*}_{P1_Y}$ = 0.16 m$_e$ for P1 and m$^{*}_{P2_Z}$ = 0.25 m$_e$, m$^{*}_{P2_Y}$ = 0.17 m$_e$ for P2.  Such closely spaced multibands near the VBM are beneficial for improving the ZT coefficient.  The overall ZT along the c axis direction (i.e. armchair direction) is calculated by
     ZT$_{max}$  $\propto$  [$\gamma_1$ $\tau_1$ $\sqrt{m^*_{X_1} m^*_{Y_1} / m^*_{Z_1}}$ + $\gamma_2$ $\tau_2$ $\sqrt{m^*_{X_2} m^*_{Y_2} / m^*_{Z_2}}$+......]
 where m$^{*}_{X_i/Y_i/Z_i}$ is the effective mass for band i of the charge carriers along the a/b/c axis, $\gamma_i$ is the degeneracy of the bands (two for each band discussed here), and $\tau_i$ is the relaxation time of the carriers along the transport direction for band i \cite{Rowe1995CRC}.  The highly anisotropic thermoelectric coefficiency can be explained by the measured dispersions. For instance, the ZT coefficient along the a axis (interlayer) direction is related to $\sqrt{m^*_{Y} m^*_{Z} / m^*_{X}}$, which is much smaller than that along the in-plane b/c axis direction due to the weak dispersion along the $\Gamma$-X direction and the resulting large effective mass m$^{*}_{X}$.  Similarly, the slightly smaller m$^{*}_{Y}$ compared to m$^{*}_{Z}$ leads to a slightly larger ZT along the b axis direction, which is in good agreement with the measured ZT \cite{Zhao2014Ultralow,Zhao2016Ultrahigh}.  Besides the effective mass, another key contributing factor for its high ZT is the intervalley scattering in the almost degenerate band structure.  Namely, the in-plane dipole field can lead to a much longer relaxation time $\tau_i$, but with negligible effect on the mobility or ZT.  The more complex band structure of layered SnSe with multivalleys and a larger effective mass give rise to a higer ZT than that of rock salt structured PbTe and SnTe \cite{Heremans2008Enhancement}.

We further discuss the effect of hole and electron doping on the thermoelectric performance. We note that ZT $\propto$ S$^2$ and the Seeback coeffeicient S is determined by the density of state (DOS), average effective mass m$^{*}_d$ =(m$^{*}_{X}$m$^{*}_{Y}$m$^{*}_{Z}$)$^{1/3}$$\gamma$$^{2/3}$, and carrier concentration n by S $\propto$ m$^{*}_d$ n$^{-2/3}$  \cite{Pei2011Convergence,Pei2012Band,Heremans2012Resonant}.  Hole doping shifts  these multivalleys toward the Fermi energy  and increases their contributions.  The decrease of S caused by the increase of carrier concentration n is therefore balanced by the enhanced contribution of multivalleys with a large m$^{*}_d$.

In addition to hole doping, further investigation on the thermoelectric performance related to the conduction bands of heavily electron doped SnSe is also critical.  Since the conduction bands of our semiconducting SnSe sample are only slighly above E$_F$, we can easily shift them to below E$_F$ through potassium (K) deposition (Fig.~\ref{Fig3}(a)) to investigate the effect of electron doping. The amount of K deposited can be estimated from the  number of peaks in the core level spectra of K 3p (Fig.~\ref{Fig3}(b)). The first K 3p core level peak emerges at about 18 eV in binding energy, and grows in intensity as K atoms are deposited on the sample surface.  As the second K over layer starts to grow, a pair of chemically shifted K 3p peaks appears at a higher binding energy about 18.5 eV \cite{Lundgren1993Layer, Kim2014Superconductivity}. Figure \ref{Fig3}(c-j) shows the band structure evolution of SnSe along the $\Gamma$-Y direction upon K deposition. With electron doping, two pockets emerge near E$_F$, one directly above the hole pocket P3 and the other at the $\Gamma$ point. In the initial stage of doping, the band structure does not change significantly compared to the band dispersion of bulk SnSe (see Supplemental Material SI III \cite{Supplemental} for calculated band dispersion of bulk SnSe). Comparison of the experimental data with band structure calculation (red solid lines in (c)) shows a good agreement, suggesting that these pockets are from the conduction bands.  The sharp single-valley bands near E$_F$ with an effective mass of m$^*$ = 0.22 m$_e$ indicates that highly electron doping cannot enhance the thermoelectric performance as much as hole doping.

More importantly, high electron doping can lead to a significant change of the band gap.  From panel (b) to (j), the pocket at the $\Gamma$ point near E$_F$ becomes larger with increasing K dosage and shows a parabolic dispersion. This parabolic band is the renormalized conduction band and is not caused by potassium-related quantum well states (QWS) as reported in other samples \cite{Alidoust2014Observation,Kim2014Superconductivity} for the following reasons. Firstly, this pocket is in good agreement with the calculated conduction band (Fig.~\ref{Fig3}(c) and Fig.~\ref{Fig4}(a)), suggesting that it is the conduction band. Secondly, QWS emerge suddenly without shifting in energy \cite{Alidoust2014Observation}, while this pocket is originated from the conduction band and evolves  continuously, i.e. graduately shift in energy). Moreover, two QWS are expected when the thickness of K reaches 2 ML \cite{Kim2014Superconductivity}, while we still observe only one band centered at $\Gamma$ (Fig.~3(h-j)). Thirdly, upon K deposition, the valence band first shifts down in energy and then moves up gradually (see arrows in Fig.~4(a-n) and schematic drawing in Fig.~4(p)). Such anomalous shift of the valence bands to lower binding energies with increasing electron doping indicates an external field on the sample to renormalized its band structure, as a result of electron-electron interaction, i.e. the negative electronic compressibility effect \cite{He2015Spectroscopic,Riley2015Negative}. As the conduction band shifts to lower energy, a pocket emerges from the valence band directly below and it shifts in opposite direction to the conduction band. This leads to a significant reduction of the band gap.

The change of the gap size is also clearly observed in data taken along the $\Gamma$-Z direction (Fig.~\ref{Fig4}(a-n)), where the CBM and VBM shift toward each other,  resulting in a reduced band gap.  Figure \ref{Fig4}(p) summarizes the evolution of the band gap as a function of K dosage, where the band gap is widely tunable from 1.2 eV to 0.4 eV. A schematic drawing of the band evolution is shown in (q). The tunable electronic state of SnSe with continuous electron doping is likely to be induced by a vertical electric field from K, which affects the real-space distribution of valence band (VB) and conduction band (CB) electrons. By shifting the VB and CB in opposite directions, the vertical electric field can lead to a renormalization of the band structure with a reduced band gap \cite{Khoo2004Tuning,Zheng2008Scaling,Ramasubramaniam2011Tunable,Yue2012Bandgap}. A reduction of band gap size by surface doping \cite{Kim2015ChemInform} and electric field \cite{Li2014Modulation} has been reported in black phosphorous with a similar crystal structure. Although the exact physical mechanism still awaits further theoretical investigations, our work provides an important opportunity to tune the band gap of layered SnSe and suggests that such tunability may be extended to other layered monochalcogenides with similar crystal structure, providing new possibilities in the design and optimization of electronic and optoelectronic devices.

\section{CONCLUSION}

To summarize, by performing a systematic ARPES study of semiconducting SnSe single crystals, we reveal the electronic origin of the excellent thermoelectric properties upon hole doping. We also find the electronic structure of SnSe is widely tunable via surface doping of K on SnSe. These results provide new opportunities for designing and optimization of electronic and optoelectronic devices, and pave the way for studying a number of similar group IV\text{-}VI materials with interesting electronic properties and wide band gap tunability.

\section{ACKNOWLEDGEMENTS}
This work is supported by the National Natural Science Foundation of China (Grant No.~11725418, 11334006, ), Ministry of Science and Technology of China (Grant No.~2016YFA0301004 and 2015CB921001) and Science Challenge Project (No. 20164500122).

\bibliography{reference}

\end{document}